\documentclass[aps,prb,showpacs,twocolumn]{revtex4}
\usepackage{graphicx}
\usepackage{epsfig}

\def  \bsig    {\mbox{\boldmath$\sigma$}}

\def  \btau  {\mbox{\boldmath$\tau$}}

\usepackage{colordvi}
\usepackage{color}

\begin{document}

\title{Spin dephasing and pumping in graphene due to random spin-orbit interaction}
\author {V. K. Dugaev$^{1,2}$, E. Ya. Sherman$^{3,4}$, and J. Barna\'s$^{5}$\footnote{Also at Faculty of Physics,
Adam Mickiewicz University, ul. Umultowska 85, 61-614 Pozna\'n,
Poland}}

\affiliation{$^1$Department of Physics, Rzesz\'ow University of Technology,
al. Powsta\'nc\'ow Warszawy 6, 35-959 Rzesz\'ow, Poland \\
$^2$Department of Physics and CFIF, Instituto Superior T\'ecnico, TU Lisbon,
Av. Rovisco Pais, 1049-001 Lisbon, Portugal \\
$^3$Department of Physical Chemistry, Universidad del Pa\'is Vasco UPV-EHU,
48080 Bilbao, Spain \\
$^4$ IKERBASQUE, Basque Foundation for Science, 48011, Bilbao, Spain  \\
$^5$Institute of Molecular Physics, Polish Academy of Sciences,
ul. Smoluchowskiego 17, 60-179 Pozna\'n, Poland}
\date{\today }

\begin{abstract}
We consider spin effects related to the random spin-orbit
interaction  in graphene. Such a random interaction can result
from the presence of ripples and/or other inhomogeneities at the
graphene surface. We show that the random spin-orbit interaction
generally reduces the spin dephasing (relaxation) time, even if the
interaction vanishes on average. Moreover, the random spin-orbit
coupling also allows for spin manipulation with an external
electric field. Due to the spin-flip interband as well as
intraband optical transitions, the spin density can be effectively
generated by periodic electric field in a relatively broad range
of frequencies.
\end{abstract}

\maketitle

\affiliation{$^1$Department of Physics, Rzesz\'ow University of Technology,
al. Powsta\'nc\'ow Warszawy 6, 35-959 Rzesz\'ow, Poland \newline
$^2$Department of Physics and CFIF, Instituto Superior T\'ecnico, TU Lisbon,
Av. Rovisco Pais, 1049-001 Lisbon, Portugal \newline
$^3$Department of Physical Chemistry, Universidad del Pa\'is Vasco, Bilbao,
Spain \newline
$^4$Adam Mickiewicz University, ul. Umultowska 85, 61-614 Pozna\'n, Poland
\newline
$^5$Institute of Molecular Physics, Polish Academy of Sciences, ul.
Smoluchowskiego 17, 60-179 Pozna\'n, Poland}

\section{Introduction}

Graphene is currently  attracting much attention as a new
excellent material for modern electronics
\cite{novoselov04,novoselov05,geim07}. The natural
two-dimensionality of graphene matches perfectly to the dominating
planar technology of other semiconducting materials, and
correspondingly gives the way to creating new hybrid systems.
However, the most striking properties of graphene are not directly
related to its two-dimensionality.  {Due to the
bandstructure effects, electrons in pure graphene can be described
by the relativistic Dirac Hamiltonian, leading to the linear
electron energy spectrum near the Dirac points. As a result, the
electronic and transport properties of graphene are significantly
different from those of any other metallic or semiconducting
material,\cite{geim07,neto09,peres09,rakyta} except (to some
extent) its parent material -- the clean graphite.\cite{graphite}}

It has been also suggested that graphene may have good
perspectives as a new material for applications in
spintronics.\cite{nishioka07,cho07,guimaraes10,yazyev10} The
intrinsic spin-orbit interaction in graphene is usually very
small, and therefore one can expect extremely long spin dephasing (relaxation)
time. \cite {hernando06,min06,yao07,gmitra09,nanotubes} Thus, spin injected
to graphene, for instance from  ferromagnetic contacts, can
maintain its coherence for a relatively long time.
Experiments demonstrate spin relaxation times for various 
graphene-based systems spanned over several orders of magnitude\cite{tombros07,tombros08,maassen,yang,han}
with some of them being much shorter than expected.\cite{tombros07,tombros08} The reason of this contradiction
is not quite clear, and several different explanations
of these observations have been already put forward.\cite{hernando07,ertler09,zhou10}
In this paper we present another
model based on the  random Rashba spin-orbit interaction.\cite{sherman03,glazov05}
Physical origin of such random spin-orbit
interaction can be related to the ripples existing at the surface
of graphene \cite{meyer07,ishigami07,wehling08} and/or to some
impurities adsorbed at the surface, which randomly enhance the
magnitude of spin-orbit coupling as compared to that in the clean
graphene.\cite{ertler09}

One of the key issues in  spintronics (including graphene-based
spintronics) is the possibility of spin manipulation with an
external electric/optical field. This includes spin generation,
spin rotation, spin switching, etc. Here we consider the
possibility of spin pumping in graphene using the idea of combined
resonance in systems with Rashba spin-orbit
interaction.\cite{rashba61,rashba03} The possibility of spin
manipulation using optical excitation\cite{zaharchenya} is based
on various mechanisms of spin-orbit interaction in semiconductor systems.
In particular, the spin polarization appears in systems with
regular spin-orbit coupling, subject to periodic electric field.
\cite{tarasenko06,raichev07,khomitsky} It has been shown recently, that the
random spin-orbit interaction also can be applied to generate spin
polarization in symmetric semiconductor quantum wells.
\cite{dugaev09} In this paper we show that similar method can be
used to generate spin polarization in graphene with random Rashba
spin-orbit interaction. To do this we analyze the intensity of
optically-induced spin-flip transitions assuming two-dimensional
massless Dirac model of electron energy spectrum in graphene, and
calculate the magnitude of spin-polarization induced by the optical
pumping.

The paper is structured as follows. In section 2 we describe the
model Hamiltonian assumed for graphene. Spin dephasing due to the
random spin-orbit Rashba coupling is calculated in section 3. In turn,
spin pumping by an external electric field is considered in
section 4. Final conclusions are presented in section 5.

\section{Model}

To describe electrons and holes in the vicinity of Dirac points
we use the model Hamiltonian $\mathcal{H}_0$ which is
sufficient when  considering the effects related to low-energy
electron and hole excitations. We also include the spin-dependent
perturbation in the form of a spatially fluctuating Rashba spin-orbit
interaction, $\mathcal{H}_{\rm so}$. Thus, the system Hamiltonian can
be written as (we use system of units with $\hbar\equiv 1$)

{\begin{eqnarray}
&&\mathcal{H}=\mathcal{H}_0+\mathcal{H}_{\rm so}, \\
&&\mathcal{H}_0=v\, \btau \cdot \mathbf{k}, \label{H01}\\
&&\mathcal{H}_{\rm so}=\frac{\lambda (\mathbf{r})}{v}
\,\left(\frac{\partial\mathcal{H}_0}{\partial k_x}\sigma _y-\frac{\partial\mathcal{H}_0}{\partial k_y}\sigma_x\right) \nonumber\\
&&=\lambda (\mathbf{r})\, \left( \tau _x\sigma _y-\tau
_y\sigma _x\right) ,  \label{Hrso}
\end{eqnarray}
where $v$ is the electron velocity, $\lambda (\mathbf{r})$ is the
random spin-orbit parameter, $\mathbf{r}=(x,y)$ is the
two-dimensional coordinate, and $\btau $ and $\bsig $ are the
Pauli matrices acting in the sublattice and spin spaces,
respectively.\cite{kane05} Equations (\ref{H01}) and (\ref{Hrso})
show that spin-orbit coupling can be described by a conventional
Rashba Hamiltonian, proportional to  $v_x\sigma_y - v_y\sigma _x$,
where the velocity components $v_x,v_y$ are, in general, obtained
with the unperturbed Hamiltonian $\mathcal{H}_0$.} It is well
known that there is an intrinsic (internal) spin-orbit coupling in
graphene, which is related to relativistic corrections to the
crystal field of the corresponding lattice. { In
addition, a spatially uniform  Rashba field can be induced by the
substrate on which the graphene sheet is located. The reported
results suggest that these interactions, which can be considered
as independent sources of spin relaxation, are either very weak
\cite{gmitra09,gold} or do not contribute to the dephasing rate
by symmetry reasons.\cite{zhou10} Therefore, we will neglect them
in our considerations and will briefly discuss their role in the
following.}

The Schr\"{o}dinger equation, $(\mathcal{H}_{0}-\varepsilon
)\,\psi _{\mathbf{k}}=0$, for the pseudospinor components of the
wavefunction $\psi _{\mathbf{k}}$ is
\begin{equation}
\left(
\begin{array}{cc}
-\varepsilon  & vk_{-} \\
vk_{+} & -\varepsilon
\end{array}
\right) \left(
\begin{array}{c}
\varphi _{\mathbf{k}} \\
\chi _{\mathbf{k}}
\end{array}
\right) =0,  \label{eq:8}
\end{equation}
where $k_{\pm }=k_{x}\pm ik_{y}$. The normalized solutions
corresponding to the eigenstates $\varepsilon _k=\pm vk$ of
Hamiltonian $\mathcal{H}_{0}$  can be written in the form
\begin{equation}
\psi _{\mathbf{k}\sigma \pm }(\mathbf{r})
=\frac{e^{i\mathbf{k}\cdot \mathbf{r}}}
{\sqrt{2}}\left( \left| 1\sigma \right\rangle \pm \frac{k_{+}}{k}\left| 2\sigma
\right\rangle \right) ,
\label{eq:9}
\end{equation}
where the $\pm $ signs correspond to the states in upper and lower
branches, respectively.

We assume that the average value of spin-orbit interaction vanishes, while
the spatial fluctuation of $\lambda (\mathbf{r})$ can be described
by the correlation function $F\left( \mathbf{r-r}^{\prime }\right)$ of a certain form,
\begin{eqnarray}
&&\left\langle \lambda (\mathbf{r})\right\rangle =0, \\
&&C(\mathbf{r}-\mathbf{r^{\prime }})\equiv \langle \lambda (\mathbf{r}%
)\,\lambda (\mathbf{r}^{\prime })\rangle =\left\langle \lambda
^{2}\right\rangle F\left( \mathbf{r-r}^{\prime }\right) .  \label{eq:5}
\end{eqnarray}

When calculating spin dephasing, one can consider only the
electron states corresponding to the upper branch (conduction
band) of the energy spectrum, $\varepsilon _{k}=vk$. The
spin-flip scattering from the random potential determines the spin
relaxation in this particular band, while from symmetry of the
system follows that spin dephasing in the lower (valence) band is
the same. The intraband matrix elements of the random spin-orbit
interaction (\ref{Hrso}) in the basis of wavefunctions
(\ref{eq:9}) for the conduction band form the following matrix in
the spin subspace
\begin{equation}
V_{\mathbf{kk^{\prime }}}=\lambda _{\mathbf{kk^{\prime }}}\left(
\begin{array}{cc}
0 & -ik_{-}/k \\
ik_{+}^{\prime }/k^{\prime } & 0
\end{array}
\right),  \label{Vkk}
\end{equation}
where $\lambda _{\mathbf{kk^{\prime }}}$ is the Fourier component
of the random spin-orbit coupling. Since scattering from the random spin-orbit
potential is elastic, only the intraband transitions contribute to
the spin relaxation.

\section{Spin dephasing}

To demonstrate how the random spin-orbit coupling works in
graphene and how its effects can be observed in experiment, as
well as to compare graphene and conventional two-dimensional
semiconductor structures, we calculate in this Section the
corresponding spin dephasing time. For this purpose we use the
kinetic equation for the density matrix (Wigner distribution
function), \cite{tarasenko06_2,dugaev09}
\begin{equation}
\frac{\partial \rho _{\mathbf{k}}}{\partial t}=\mathrm{St}\,\rho _{\mathbf{k}}.
\end{equation}
The collision integral $\mathrm{St}\,\rho _{\mathbf{k}}$ on the
right-hand side of this equation is due to the spin-flip
scattering from the random spin-orbit interaction,
\begin{eqnarray}
\mathrm{St}\,\rho _{\mathbf{k}} &=&\pi \sum_{\mathbf{k^{\prime }}}\left( 2V_{%
\mathbf{kk^{\prime }}}\rho _{\mathbf{k^{\prime }}}V_{\mathbf{k^{\prime }k}%
}-V_{\mathbf{kk^{\prime }}}V_{\mathbf{k^{\prime }k}}\rho _{\mathbf{k}}-\rho
_{\mathbf{k}}V_{\mathbf{kk^{\prime }}}V_{\mathbf{k^{\prime }k}}\right)
\nonumber \\
&&\times \delta \left( \varepsilon _{\mathbf{k}}-\varepsilon _{\mathbf{%
k^{\prime }}}\right) .  \label{ST}
\end{eqnarray}
We assume the following form of the density matrix:
\begin{equation}
\rho _{\mathbf{k}}=\rho_{0k}+s_{\mathbf{k}}\sigma _{z},  \label{rho}
\end{equation}
where the first term $\rho_{0k}$ corresponds to the
spin-unpolarized equilibrium state. On substituting (\ref{Vkk})
and (\ref{rho}) into Eq.~(\ref{ST}) we find
\begin{equation}
\mathrm{St}\,\rho _{\mathbf{k}}=-2\pi \sigma _{z}\sum_{k^{\prime }}C(\mathbf{%
q})\left( s_{\mathbf{k}}+s_{\mathbf{k^{\prime }}}\right) \,\delta \left(
\varepsilon _{\mathbf{k}}-\varepsilon _{\mathbf{k^{\prime }}}\right),
\label{ST1}
\end{equation}
where $C(\mathbf{q})\equiv\left\langle\lambda_{\mathbf{kk^{\prime
}}}^{2}\right\rangle$ and $\mathbf{q}=\mathbf{k^{\prime }-k}$.
Assuming that $s_{\mathbf{k}}$ does not depend on the point at the
Fermi surface we obtain
\begin{equation}
\mathrm{St}\,\rho_{\mathbf{k}}
=-\frac{4k\sigma _{z}s_{k}}{\pi v}\int_{0}^{2k}\frac{C(\mathbf{q})}{\sqrt{%
4k^{2}-q^{2}}}dq.
\label{ST2}
\end{equation}

For definiteness, we assume that the characteristic spatial range
of the random spin-orbit fluctuations is $R$, and take $C(\mathbf{q})$ in
the following form:
\begin{equation}
C(\mathbf{q})=2\pi \langle \lambda ^{2}\rangle R^{2}e^{-qR},
\end{equation}
{satisfying the normalization condition
\begin{equation}
\int C(\mathbf{q})\frac{d^2q}{(2\pi)^2}=\langle \lambda ^{2}\rangle.
\end{equation}
The resulting spin relaxation rate is not strongly sensitive to
the shape of the correlator. However, the applicability of the
approach based on Eq.(\ref{ST1}) depends on the ratio of electron
mean  free path $\ell$ to $R$, being valid only if $\ell/R\gg1$.
In such a case, typically realized in graphene, the electron spin
experiences indeed random, weakly correlated in time fluctuations
of the spin-orbit coupling. In addition, we can safely neglect the
effect of the random spin-orbit coupling on the momentum
relaxation rate.} Finally, for the  spin dephasing time we obtain
the following expression:
\begin{eqnarray}
\frac{1}{\tau _{k}^{s}} &=&\frac{8k}{v}\left\langle \lambda
^{2}\right\rangle R^{2}\int_{0}^{2kR}\frac{e^{-x}dx}{\sqrt{4k^{2}R^{2}-x^{2}}%
}  \nonumber \\
&=&\frac{4\pi k}{v}\left\langle \lambda ^{2}\right\rangle R^{2}\left[
I_{0}(2kR)-L_{0}(2kR)\right] ,  \label{tausk1}
\end{eqnarray}
where $I_{0}(x)$ and $L_{0}(x)$ are the modified Bessel and Struve functions
of zeroth order, respectively.

\begin{figure}
%\vspace*{-1cm}
%\hspace*{-0.5cm}
\includegraphics*[scale=0.6]{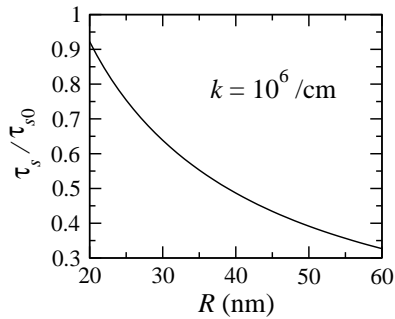}
%\vspace*{-1.5cm}
%\hspace*{-0.5cm}
\caption{Spin dephasing time as a function of the characteristic
range $R$ of the random spin-orbit fluctuations.}
\end{figure}

In the limiting  semiclassical ($kR\gg 1$) and quantum ($kR\ll1$) cases we find
\begin{equation}
\frac{1}{\tau _{k}^{s}}\simeq \frac{4R}{v}\left\langle \lambda
^{2}\right\rangle \left\{
\begin{array}{cc}
1, & kR\gg 1, \\
\\
\pi kR, & kR\ll 1.
\end{array}
\right.
\label{tausk2}
\end{equation}
For $kR\gg 1$, the result in Eq.(\ref{tausk2}) can be interpreted
as the special realization of the Dyakonov-Perel' spin relaxation mechanism. To see this, we
note that the electron spin  rotates at the rate $\Omega \sim
\left\langle \lambda ^{2}\right\rangle ^{1/2}$, with the
precession direction changing randomly at the timescale of the
order of time that electron needs to pass through one domain of
the size $R$, i.e., $\tau _{R}\sim R/v.$ The resulting spin
relaxation rate $1/\tau _{k}^{s}$ is of
the order of $\Omega^{2}\tau_{R}.$ It is worth mentioning
that at given spatial and energy scale of the fluctuating 
spin-orbit field, the decrease in the electron free path 
and in the momentum relaxation time leads to the decrease in the spin dephasing
rate: if $\ell\ll R$, electron spin interacts with the local rather
than with the rapidly changing random field, and spin relaxation
rate becomes of the order of $\left<\lambda^{2}\right>\tau$, where $\tau$ is the momentum relaxation time. 
This agrees qualitatively with the observations of Ref.[\onlinecite{yang}],
however, a quantitative comparison needs a more detailed analysis. 

Taking examples with typical values 
$v=10^{8}$ cm/s, $R\sim 50$ nm,  and $\langle\lambda^{2}\rangle\sim 500$ $\mu$eV$^{2}$,
similar to what can be expected from
Ref.[\onlinecite{hernando06}],  we obtain $\tau _{k}^{s}$ less than or of the order of 10 ns.
As one can see from Eq.(\ref{tausk2}), the spin
relaxation for  small $kR\ll1 $ is suppressed, as can be
understood in terms of the averaging of the random field over a
large $1/k^{2}\gg R^{2}$ area. {The full
$k$-dependence in Eq.(\ref{tausk2}) implies that the spin
dephasing rate is proportional to $n^{1/2}$ at low carrier
concentrations $n$ and is independent of $n$ at higher ones.}
Therefore, at the charge neutrality point, where $n=0$, the spin relaxation
vanishes, in agreement with the observations of Ref.[\onlinecite{han}].
Moreover, our approach qualitatively agrees with the increase in the
spin relaxation time in the bilayer graphene compared to the single layer 
one \cite{han}: the transverse rigidity of the bilayer 
can be larger, thus suppressing formation of the long-range ripples, and, as 
a result, the random spin-orbit coupling.

Equation (\ref{tausk2}) shows that, as far
as the spin dephasing is considered, the only difference between
graphene and conventional semiconductors
\cite{sherman03,glazov05,dugaev09} is related to the fact that the
electron velocity is constant for the former case and is
proportional to the momentum in the latter one. As we will see in
the next Section, this difference becomes crucial for the spin
pumping processes.

The dependence of spin relaxation time, calculated from
Eq.~(\ref{tausk1}) as a function of the characteristic domain size
$R$ of the random spin-orbit interaction is presented in Fig.~1,
were $\tau_{s0}$ is defined as $\tau_{s0}^{-1}\equiv 8\left\langle
\lambda ^{2}\right\rangle /vk$. The curves corresponding to
different values of $k>10^{6}$~cm$^{-1}$ would be practically
indistinguishable in this figure.

{Here several comments on the numerical values of
spin relaxation are in order. The values observed in experiments
on spin injection from ferromagnetic
contacts\cite{tombros07,tombros08} are of the order of 10$^{-10}$
s, two orders of magnitude less than our estimate which does not
take into account explicitly the role of the Si-based substrates.
The effect of the SiO$_2$ substrate, including the contributions
from impurities and electron-phonon coupling, was thoroughly
analyzed in Ref.[\onlinecite{ertler09}]. However, the obtained
dephasing rates were well below the experimental values and also
below the estimate obtained here, leading the authors of
Ref.[\onlinecite{ertler09}] to the suggestion of an important role
of heavy adatoms in the spin-orbit coupling.\cite{adatoms} On
the other hand, it was shown that the spin dephasing rate can be
strongly influenced at relatively high temperatures  by the
electron-electron collisions.\cite{zhou10} However, including
these collisions does not bring theoretical values closer to the
experimental ones. 

The discrepancies between  theory and
experiment and between experimental data obtained  
on different systems call for a more detailed analysis of the experimental
situation, including the dependence of spin relaxation time on the
device functional properties and the experimental techniques
applied.}

\section{Spin pumping}

Let us consider now  spin pumping by an external electromagnetic
periodic field corresponding to the vector potential
$\mathbf{A}(t)$. We assume that the system described by
Eqs.~(1)-(3) is additionally in a constant magnetic field  $\mathbf{B}$. For
simplicity, we consider the Voigt geometry with the field
in the graphene plane, so that the effects of Landau quantization
are absent. Thus, the Hamiltonian $\mathcal{H}_{0}$ includes now
the constant field and can be written as
\begin{equation}
\mathcal{H}_{0}=v\,\btau \cdot \mathbf{k}+\Delta \sigma _{x},  \label{H0}
\end{equation}
where $2\Delta =g\mu_{B}B$ is the spin splitting and the
magnetic field is oriented along the $x-$axis (the electron Land\'{e}
factor for graphene is $g=2$). {The induced
spin polarization is opposite to the direction of the magnetic field.}

Since we are interested in real transitions in which the energy is
conserved, we consider interaction with a single component of
the periodic electromagnetic field,
$\mathbf{A}(t)=\mathbf{A}e^{-i\omega t},$ which enters in the
gauge-invariant form

{\begin{equation}
\mathcal{H}_{A}=-\frac{e}{c}\,\frac{\partial\mathcal{H}}{\partial \mathbf{k}} \cdot \mathbf{A}(t)=
-\frac{ev}{c}\,\btau \cdot \mathbf{A}(t),  \label{eq:18}
\end{equation}}
and in the following we treat the term $\mathcal{H}_{A}$ as a
small perturbation.

The absorption in a  periodic field (probability of field-induced
transitions in unit time) can be written as
\begin{equation}  \label{eq:19}
I(\omega )=\mathrm{Re\, Tr} \sum _{\mathbf{k}}\int \frac{d\varepsilon}{2\pi }%
\; \mathcal{H}_A\; G_{\mathbf{k}}(\varepsilon +\omega )\; \mathcal{H}_A\; G_{%
\mathbf{k}}(\varepsilon),
\end{equation}
where $G_{\mathbf{k}}(\varepsilon )$ is the Green function. In the
absence of spin-orbit interaction, the absorption (\ref{eq:19}) does not include
any spin-flip transitions. We can account for the spin-orbit interaction
(\ref{Hrso}) in the second order perturbation theory, including the
corresponding matrix elements as shown in Fig.~2. This means that
we do not consider perturbation terms as the self-energy
within a single Green function assuming that they are  already included in the electron
relaxation rate.

\begin{figure}
%\vspace*{-1cm}
%\hspace*{-0.5cm}
\includegraphics*[scale=0.6]{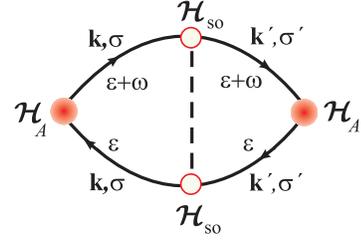}
%\vspace*{-0.5cm}
%\hspace*{-0.5cm}
\caption{(Color online.)  Feynman diagrams for the light absorption in the second
order perturbation theory with respect to the  random spin-orbit  interaction.
The coupling and electromagnetic vertices are shown as white and filled circles, respectively.}
\end{figure}

Thus, in the second order perturbation theory with respect  to the
random spin-orbit interaction $\mathcal{H}_{\rm so}$ we obtain
\begin{eqnarray}
I(\omega ) &=&\mathrm{Re\,Tr}\,\sum_{\mathbf{kk^{\prime }}}\int \frac{%
d\varepsilon }{2\pi }\;\mathcal{H}_{A}\;G_{\mathbf{k}}^{0}(\varepsilon
+\omega )\;\mathcal{H}_{\rm so}^{\mathbf{kk^{\prime }}}\,G_{\mathbf{k^{\prime }}%
}^{0}(\varepsilon +\omega )\;  \nonumber  \label{eq:20} \\
&&\times \mathcal{H}_{A}\;G_{\mathbf{k^{\prime }}}^{0}(\varepsilon )\;%
\mathcal{H}_{\rm so}^{\mathbf{k^{\prime }k}}\;G_{\mathbf{k}}^{0}(\varepsilon ),
\end{eqnarray}
where Green's function $G_{\mathbf{k}}^{0}(\varepsilon )=\mathrm{diag}%
\left\{ G_{\mathbf{k}\uparrow }^{0}(\varepsilon
),\,G_{\mathbf{k}\downarrow }^{0}(\varepsilon )\right\} $
corresponds to the Hamiltonian $\mathcal{H}_{0} $ of
Eq.~(\ref{H0}),
\begin{equation}
G_{\mathbf{k}\sigma }^{0}(\varepsilon )=\frac{\varepsilon
+v\,\btau \cdot \mathbf{k}+\Delta \sigma +\mu }{\left( \varepsilon
-\varepsilon _{1k\sigma }+\mu +i\delta \,\mathrm{sgn}\,\varepsilon
\right) \left( \varepsilon -\varepsilon _{2k\sigma }+\mu +i\delta
\,\mathrm{sgn}\,\varepsilon \right) }, \label{Gk}
\end{equation}
with $\sigma = \pm 1$ corresponding to the spins oriented along
and opposite to the $x-$axis, respectively,
$\varepsilon _{(1,2)k\sigma }=\pm v\,k+\Delta \sigma $, 
$\delta$ being the half of the momentum relaxation rate, $\delta =1/2\tau $,
and $\mu$ denoting the chemical potential.

\begin{figure}
%\vspace*{-0.5cm}
%\hspace*{-0.5cm}
\includegraphics*[scale=0.6]{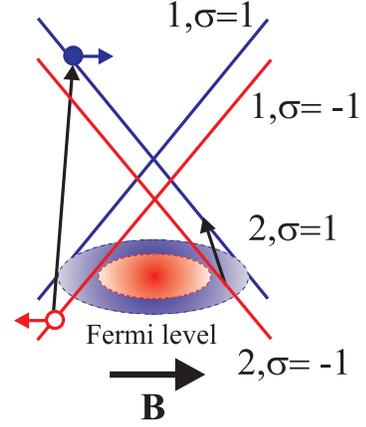}
%\vspace*{-1cm}
%\hspace*{-0.5cm}
\caption{(Color online.)  The energy spectrum and indirect spin-flip transitions
in graphene in a uniform magnetic field $\mathbf{B}$.
Long and short arrows correspond to the interband and intraband transitions,
respectively.}
\end{figure}

Diagrams in Fig. 2 show the qualitative difference between the
graphene and semiconductor quantum well with respect to the
effects of random spin-orbit coupling. In semiconductors, the external
electric field is explicitly coupled to the anomalous
spin-dependent term in the electron velocity, which is random, and
therefore the diagrams describing the corresponding transitions
include only two Green functions. In graphene, due to the absence
of randomness-originated term in the Hamiltonian
$\mathcal{H}_{A},$ four Green functions are required to take into
account the random contribution of the spin-orbit coupling. This
situation, in some sense, is more close to what is observed in the
conventional kinetic theory of normal metal conductivity, where
the coupling to the external field does not depend on the
randomness explicitly, and the additional disorder effects appear
due to the self-energy and/or due to the vertex corrections, as
in  Fig.~2.

{Upon calculating contributions from the diagrams
of Fig.~2, one finds the total rate of spin-flip and
spin-conserving transitions due to the random spin-orbit coupling
in the form
\begin{eqnarray}
I_{\rm rso}(\omega ) &=&\frac{e^{2}A^{2}}{c^{2}}\,\mathrm{Re\,Tr}\sum_{\sigma\sigma^{\prime}}
\sum_{\mathbf{kk^{\prime }}}v^{2}C(\mathbf{q})\int \frac{d\varepsilon
}{2\pi }(\btau \cdot \mathbf{n}_{A})  \nonumber  \label{eq:22} \\
&&\times G_{\mathbf{k}\sigma }^{0}(\varepsilon +\omega )\tau_{-}
G_{\mathbf{k^{\prime }}\sigma
^{\prime }}^{0}(\varepsilon +\omega )(\btau \cdot\mathbf{n}_{A})  \nonumber \\
&&\times G_{\mathbf{k^{\prime }}\sigma ^{\prime }}^{0}(\varepsilon )\,
\tau _{+}G_{\mathbf{k}\sigma
}^{0}(\varepsilon ),
\label{Irso}
\end{eqnarray}
where $\tau _{\pm }=\tau _{_{x}}\pm i\tau _{y},$ and
$\mathbf{n}_{A}$ describes the direction of $\mathbf{A}$.}  In the
following we assume linear polarization of light,
$\mathbf{A}=(A,\,0)$. After calculating the trace and integrating
over energy $\varepsilon $ in Eq.~(\ref{Irso}) one finds a rather
cumbersome expression (see Appendix) consisting of several terms,
each of them corresponding to transitions between certain branches
of the spectrum (Fig.~3). For definiteness, we locate the chemical
potential in the valence band. Correspondingly, only the
transitions  from the bands $(2\,\uparrow)$ and $(2\,\downarrow)$
to the unoccupied states in the bands $(1\,\uparrow)$,
$(1\,\downarrow)$, $(2\,\uparrow)$ and $(2\,\downarrow)$ are
possible. We will concentrate on the optically induced spin-flip
transitions contributing to the optically-generated spin pumping.
Hence, we do not consider spin-conserved transitions contributing
to the usual Drude conductivity.\cite{Kuzmenko,optics,ESR}

\subsection{Interband spin-flip transitions}

\begin{figure}
%\vspace*{-1cm}
%\hspace*{-0.5cm}
\includegraphics*[scale=0.6]{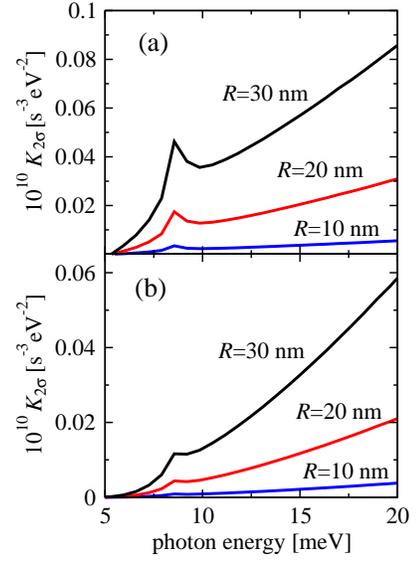}
%\vspace*{-1.5cm}
%\hspace*{-0.5cm}
\caption{(Color online.) The parameter $K_{2\downarrow}(\omega)$ for the rate
of the spin-flip interband transitions in the high-frequency domain  (a) and
$K_{2\uparrow}$ (b) for different values
of the range parameter $R$ describing characteristic size of the
fluctuations in spin-orbit interaction and $\langle\lambda^{2}\rangle=100$ $\mu$eV$^{2}$.}
\end{figure}

Let us consider first the spin-flip transitions from the valence
to conduction bands,
such as $\overline{\mathbf{k}}_{2}\rightarrow $ $\overline{\mathbf{k}}%
_{1}^{\prime }.$ For convenience we introduce the parameter $\overline{%
\mathbf{k}}_{j}=\left( \mathbf{k,\sigma }_{x}\right)_{j},$
describing the momentum and spin projection for an electron in the
subband $j.$ The corresponding expression for the transition
rate can be obtained from the equations presented in the Appendix,
and considerably simplified by taking
into account that: (i) in the nonsingular terms $\varepsilon _{\overline{\mathbf{k}}%
_{1}}-\varepsilon _{\overline{\mathbf{k}}_{2}}$ can be substituted by $%
\omega ,$  (ii) in the semiclassical limit $kR\gg 1$ the energy
change due to the change in the momentum is small compared to
$1/\tau$, and (iii) the photon energy is much larger than the
characteristic low-energy scale parameters, i.e., $\omega \gg
1/\tau ,$ and $\omega \gg \Delta$. We mention that the linear in
$\Delta$ terms have to be kept despite $\Delta\ll\omega$ since the
resulting spin pumping rate, determined by the contributions of
both initial spin states, is linear in $\Delta$. 
{The expressions for the parameters $K_{2\sigma}(\omega)$, which
determine the spin-flip rate as introduced in the Appendix,
Eq.(\ref{Ijsigma}), can be then simplified, and as a result one
obtains the following formula for the spin-flip rate in the
relevant frequency domain:
\begin{eqnarray}
I_{2\sigma \to 1\sigma ^{\prime}}(\omega)
=4\pi\sigma \frac{e^{2}A^{2}}{c^{2}}
\frac{v^{4}}{\omega }\sum_{\mathbf{kk^{\prime }}}C\left(
\mathbf{q}\right) \left[ f(\varepsilon_{\overline{\mathbf{k}}_{2}})
-f(\varepsilon_{\overline{\mathbf{k}}_{1}^{\prime }})\right]
\nonumber \\
\times \frac{k^{2}-kq\cos \varphi }
{\varepsilon _{\overline{\mathbf{k}}_{2}}+\omega
-\varepsilon _{\overline{\mathbf{k}}_{1}}}
\, \frac{\delta
\left( \varepsilon _{\overline{\mathbf{k}}_{2}}+\omega
-\varepsilon _{\overline{\mathbf{k}}_{1}^{\prime }}\right) }
{\left( \varepsilon _{\overline{\mathbf{k}}_{2}}+\omega
-\varepsilon _{\overline{\mathbf{k}}_{2}^{\prime }}\right)
\left( \varepsilon _{\overline{\mathbf{k}}_{2}}
-\varepsilon _{\overline{\mathbf{k}}_{2}^{\prime }}\right) }\, .\hskip0.2cm
\label{I21}
\end{eqnarray}}
The transitions are constrained due to the $\delta$-function in Eq.~(\ref{I21})
corresponding to the energy conservation with the change in the
momentum $\mathbf{q}$ (here we use $\varphi =\cos ^{-1}(\mathbf{k},\mathbf{q})$)
\begin{equation}
\delta \left( \varepsilon _{\overline{\mathbf{k}}_{2}}+\omega -\varepsilon _{%
\overline{\mathbf{k}}_{1}^{\prime }}\right) =\frac{|\omega -vk-2\sigma
\Delta |\;\delta (\varphi -\varphi _{0})}{v^{2}kq\,|\sin \varphi _{0}|}\,,
\label{delta}
\end{equation}
where $\varphi _{0}$ is a solution of the equation
\begin{equation}
vk+2\sigma \Delta -\omega +v\sqrt{k^{2}-2kq\cos\varphi+q^{2}}=0,
\label{argument}
\end{equation}
which gives us the condition for a minimum value of momentum in Eq.~(\ref{argument}),
$vk_{\min }=\omega -2\sigma \Delta $. The energy conservation determines the
angle $\varphi _{0}$ between the vectors $\mathbf{k}$ and $\mathbf{q}$ as
\begin{equation}
\cos \varphi _{0}=\frac{q}{2k}+\frac{\omega -2\sigma \Delta }{vq}\left( 1-%
\frac{\omega -2\sigma \Delta }{2vk}\right).
\label{cosphi}
\end{equation}
The usual condition of $|\cos \varphi _{0}|<1$ leads to the following
restrictions in the integration over $q$ in Eq.~(\ref{I21}):

\begin{figure}
%\vspace*{3cm}
%\hspace*{-0.5cm}
\includegraphics*[scale=0.6]{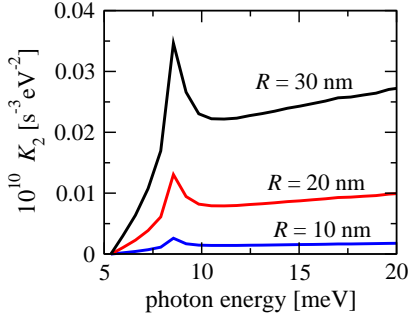}
%\vspace*{1cm}
%\hspace*{-0.5cm}
\caption{(Color online.)  The parameter $K_{2}(\omega)\equiv K_{2\downarrow}(\omega)-K_{2\uparrow}(\omega)$ for the
spin injection rate for different $R$. Other parameters are the same as Fig.4.}
\end{figure}

a) if $\omega /2-\sigma \Delta <vk<\omega -2\sigma \Delta $ then $%
vk-|vk+2\sigma \Delta -\omega |<vq<vk+|vk+2\sigma \Delta -\omega |$,

b) if $\omega /2-\sigma \Delta >vk$ then $-vk+|vk+2\sigma \Delta -\omega
|<vq<vk+|vk+2\sigma \Delta -\omega |$.

Accounting for all these conditions, the integral over $k$ and $q$
in Eq.~(\ref{I21}) can be calculated numerically.
We used the following parameters: $\mu
=-3$~meV, corresponding to the Fermi momentum $k_F\approx 5\times
10^4$ cm$^{-1}$ and hole concentration $\approx 4\times 10^8$
cm$^{-2}$, $\Delta =1$~meV, and $1/\tau=1$~meV. These parameters
correspond to a rather clean graphene and strong Zeeman splitting
of the bands (magnetic field of 17 T). For numerical accuracy, the
calculations were performed with exact Eqs.(\ref{J2down}) and (\ref{J2up}).
{ The results of
numerical calculations for the interband spin-flip transitions are
presented in Fig.~4.
Figure 5 presents the quantity describing the total spin injection rate
for the interband transitions in this frequency domain, $K_{2}(\omega)\equiv K_{2\downarrow}(\omega)-K_{2\uparrow}(\omega)$.
The positive sign of $K_{2}(\omega)$ corresponds to the fact that the
absolute value of spin density \textit{decreases} due to
the pumping.} The peaks correspond
to the absorption edge of direct optical transitions, where the
transition probability rapidly increases, and the increase at
large frequencies is due to the linear energy dependence of the
density of states.

\subsection{Intraband transitions in the hole subband}

Using the general formulas (\ref{J2down}) and (\ref{J2up}) we can write down the
expression for the spin-flip intraband transitions within the
valence band in a low-frequency region $\omega \ll \left| \mu
\right| $. Such transitions are associated with a relatively large
change of the momentum in the absorption process, and one can
expect that they give a smaller transition rate. Upon
taking into account that $\varepsilon _{\overline{\mathbf{k}}%
_{1}}-\varepsilon _{\overline{\mathbf{k}}_{2}}=-2\mu \gg \omega ,$ and $%
\left| \varepsilon _{\overline{\mathbf{k}}_{2}}-\varepsilon _{\overline{%
\mathbf{k}}_{2}^{\prime }}\right| \ll \left| \mu \right|$, we find
{
\begin{eqnarray}
I_{2\sigma \to 2\sigma ^{\prime }}(\omega )
=-4\pi\sigma \frac{
e^{2}A^{2}}{c^{2}}
\frac{v^{4}\tau^{2}}{\mu^{2}(1+\omega^{2}\tau^{2})} \nonumber \\
\times \sum_{\mathbf{kk^{\prime }}}
C\left( \mathbf{q}\right)
\left[ f(\varepsilon _{\overline{\mathbf{k}}_{2}})
-f(\varepsilon _{\overline{\mathbf{k}}_{2}^{\prime }})\right]
\nonumber \\
\times \varepsilon _{\overline{\mathbf{k}}_{2}}\varepsilon _{\overline{%
\mathbf{k}}_{2}^{\prime }}(k^{2}-kq\cos \varphi )
\, \frac{\delta \left(
\varepsilon _{\overline{\mathbf{k}}_{2}}+\omega -\varepsilon _{\overline{%
\mathbf{k}}_{2}^{\prime }}\right) }
{(\varepsilon _{\overline{\mathbf{k}}_{2}}
+\omega -\varepsilon _{\overline{\mathbf{k}}_{1}^{\prime }})
(\varepsilon _{\overline{\mathbf{k}}_{2}}
-\varepsilon _{\overline{\mathbf{k}}_{1}^{\prime }})}.
\label{I22}
\end{eqnarray}}
The $\delta$-function can be presented in the form of
Eq.(\ref{delta}), where $ \varphi _{0}$ is a solution of the
equation
\begin{equation}
vk+2\sigma \Delta -\omega -v\sqrt{k^{2}-2kq\cos \varphi +q^{2}}=0.
\label{argument1}
\end{equation}

\begin{figure}
%\vspace*{2cm}
%\hspace*{-0.5cm}
\includegraphics*[scale=0.6]{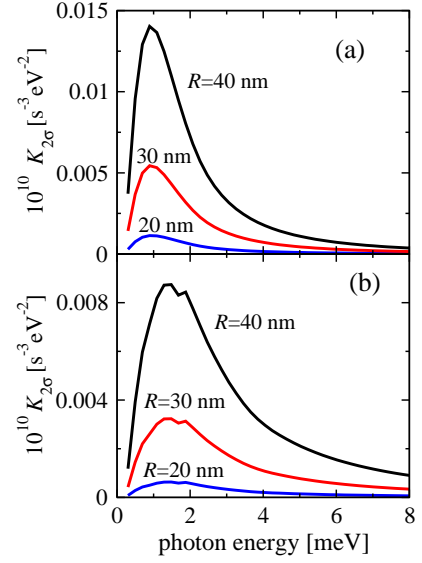}
%\vspace*{-1.5cm}
%\hspace*{-0.5cm}
\caption{(Color online.)  The parameter $K_{2\downarrow}(\omega)$ for the rate
of the spin-flip intraband transitions in the low-frequency domain  (a) and
$K_{2\uparrow}(\omega)$ (b) for different values
of the range parameter $R$. Other parameters are the same as Fig.4.}
\end{figure}

\begin{figure}
%\vspace*{6cm}
%\hspace*{-0.5cm}
\includegraphics*[scale=0.6]{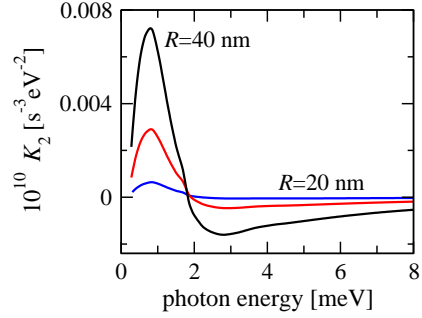}
%\vspace*{1cm}
%\hspace*{-0.5cm}
\caption{(Color online.)
The parameter $K_{2}(\omega)\equiv K_{2\downarrow}(\omega)-K_{2\uparrow}(\omega)$ for the
spin injection rate for different $R$ in the low-frequency
domain. Other parameters are the same as Fig.4.
Unmarked line corresponds to $R=30$ nm.}
\end{figure}

Equations (\ref{argument1}) and (\ref{argument}) are similar, with
the important difference in the sign in front of
$\sqrt{k^{2}-2kq\cos \varphi +q^{2}}.$ In the case of interband
transitions this corresponds to the transitions occurring at
$\omega \approx 2vk,$ while the intraband transitions occur at
lower frequencies determined by the possible momentum transfer due
to the randomness of the spin-orbit coupling. The solution exists only for
$vk>\omega -2\sigma \Delta $ and gives  Eq.(\ref{cosphi}).
However, the condition $|\cos\varphi_{0}|<1$ leads here to a
different restriction. Since only the condition $\omega -2\sigma
\Delta <2vk$ is consistent with $vk>\omega -2\sigma \Delta $, the
momentum $q$ should be then in the single range of $%
vk-|vk+2\sigma \Delta -\omega |<vq<vk+|vk+2\sigma \Delta -\omega
|$, in contrast to the case of interband transitions.

The results of calculations for the intraband transitions are
presented in Fig.~6. The intensity of such
processes is relatively small compared to the interband
transitions, and they can be seen only at low photon energies.
{Figure 7 corresponds to spin injection rate
by the intraband transitions, $K_{2}(\omega)$.}

\section{Conclusions}

We have considered certain spin effects associated with random
spin-orbit interaction in graphene. First, we have
calculated the corresponding spin relaxation time, and believe
that this mechanism can be dominating when the amplitude of
fluctuations in spin-orbit interaction is large enough. This may
happen in the presence of surface ripples with short wavelengths.
The other possibility can be related to the absorbed impurities at
both surfaces of a free-standing graphene. One can expect
especially strong random spin-orbit coupling for heavy impurity
atoms.

The second effect concerns the possibility of spin pumping by an
external electromagnetic field. The results of our calculations
show that graphene can be used as a material, in which the
electron spin density can be generated by the optical pumping. The
mechanism of pumping here is related to the spin-flip transitions
associated with the random Rashba spin-orbit interaction.

\section*{Acknowledgements}

This work is partly supported by the FCT Grant PTDC/FIS/70843/2006 in
Portugal and by the Polish Ministry of Science and Higher Education as a
research project in years 2007 -- 2010. This work of
EYS was supported by the University of Basque Country UPV/EHU grant GIU07/40,
MCI of Spain grant FIS2009-12773-C02-01, and "Grupos Consolidados UPV/EHU
del Gobierno Vasco" grant IT-472-10.

\begin{appendix}

\section{Formula for the absorption rate}

Using (\ref{Gk}) and (\ref{Irso}), after calculating the trace and
integrating over energy $\varepsilon $, we find out that spin-flip processes
can be characterized by the initial state as $I_{1\downarrow }(\omega
),I_{2\downarrow }(\omega ),I_{1\uparrow }(\omega ),$ and $I_{2\uparrow
}(\omega )$ with the corresponding transition rate  $I_{j\sigma }(\omega )$
defined as
\begin{eqnarray}
I_{j\sigma }(\omega ) &\equiv &\frac{16e^{2}A^{2}}{c^{2}}K_{j\sigma }(\omega
), \\
K_{j\sigma }(\omega ) &\equiv &-\mathrm{Im}\int v^{4}C\left( \mathbf{q}%
\right) \frac{J_{j\sigma }(\mathbf{q,}\omega )}{\omega +i\delta }\frac{%
d^{2}kd^{2}k^{\prime }}{\left( 2\pi \right) ^{4}},
\label{Ijsigma}
\end{eqnarray}
describing transitions from the band and spin states corresponding to the
subscript $j\sigma $. Here $J_{j\sigma }(\mathbf{q,}\omega )$ have the form:
\begin{widetext}
\begin{eqnarray}
J_{1\downarrow }(\mathbf{q,}\omega ) &=&\left. \left[ f(\varepsilon
_{1k\downarrow })-f(\varepsilon _{1k\downarrow }+\omega )\right] \,\frac{%
(\varepsilon _{1k\downarrow }^{2}-\Delta ^{2}+\varepsilon _{1k\downarrow
}\omega )(k^{2}-kq\cos \varphi )}{(\varepsilon _{1k\downarrow }-\varepsilon
_{1k^{\prime }\uparrow })(\varepsilon _{1k\downarrow }-\varepsilon
_{2k\downarrow })(\varepsilon _{1k\downarrow }-\varepsilon _{2k^{\prime
}\uparrow })}\right.   \nonumber  \label{eq:A1} \\
&&\times \left. \frac{1}{\left( \varepsilon _{1k\downarrow }+\omega
-\varepsilon _{1k^{\prime }\uparrow }+i\delta \right) (\varepsilon
_{1k\downarrow }+\omega -\varepsilon _{2k\downarrow }+i\delta )\left(
\varepsilon _{1k\downarrow }+\omega -\varepsilon _{2k^{\prime }\uparrow
}+i\delta \right) }\right. ,
\label{J1down}
\end{eqnarray}
\begin{eqnarray}
J_{2\downarrow }(\mathbf{q,}\omega ) &=&\left. \left[ f(\varepsilon
_{2k\downarrow })-f(\varepsilon _{2k\downarrow }+\omega )\right] \,\frac{%
(\varepsilon _{2k\downarrow }^{2}-\Delta ^{2}+\varepsilon _{2k\downarrow
}\omega )(k^{2}-kq\cos \varphi )}{(\varepsilon _{2k\downarrow }-\varepsilon
_{2k^{\prime }\uparrow })(\varepsilon _{2k\downarrow }-\varepsilon
_{1k\downarrow })(\varepsilon _{2k\downarrow }-\varepsilon _{1k^{\prime
}\uparrow })}\right.   \nonumber \\
&&\times \left. \frac{1}{\left( \varepsilon _{2k\downarrow }+\omega
-\varepsilon _{2k^{\prime }\uparrow }+i\delta \right) (\varepsilon
_{2k\downarrow }+\omega -\varepsilon _{1k\downarrow }+i\delta )\left(
\varepsilon _{2k\downarrow }+\omega -\varepsilon _{1k^{\prime }\uparrow
}+i\delta \right) }\right. ,
\label{J2down}
\end{eqnarray}
\begin{eqnarray}
J_{1\uparrow }(\mathbf{q,}\omega) &=&\left. \left[ f(\varepsilon
_{1k\uparrow })-f(\varepsilon _{1k\uparrow }+\omega )\right] \,\frac{%
(\varepsilon _{1k\uparrow }^{2}-\Delta ^{2}+\varepsilon _{1k\uparrow }\omega
)(k^{2}-kq\cos \varphi )}{(\varepsilon _{1k\uparrow }-\varepsilon
_{1k^{\prime }\downarrow })(\varepsilon _{1k\uparrow }-\varepsilon
_{2k\uparrow })(\varepsilon _{1k\uparrow }-\varepsilon _{2k^{\prime
}\downarrow })}\right.   \nonumber \\
&&\times \left. \frac{1}{(\varepsilon _{1k\uparrow }+\omega -\varepsilon
_{1k^{\prime }\downarrow }+i\delta )\left( \varepsilon _{1k\uparrow }+\omega
-\varepsilon _{2k\uparrow }+i\delta \right) (\varepsilon _{1k\uparrow
}+\omega -\varepsilon _{2k^{\prime }\downarrow }+i\delta )}\right. ,
\label{J1up}
\end{eqnarray}
\begin{eqnarray}
J_{2\uparrow }(\mathbf{q,}\omega ) &=&\left. \left[ f(\varepsilon
_{2k\uparrow })-f(\varepsilon _{2k\uparrow }+\omega )\right] \,\frac{%
(\varepsilon _{2k\uparrow }^{2}-\Delta ^{2}+\varepsilon _{2k\uparrow }\omega
)(k^{2}-kq\cos \varphi )}{(\varepsilon _{2k\uparrow }-\varepsilon
_{2k^{\prime }\downarrow })(\varepsilon _{2k\uparrow }-\varepsilon
_{1k\uparrow })(\varepsilon _{2k\uparrow }-\varepsilon _{1k^{\prime
}\downarrow })}\right.   \nonumber \\
&&\times \left. \frac{1}{(\varepsilon _{2k\uparrow }+\omega -\varepsilon
_{2k^{\prime }\downarrow }+i\delta )\left( \varepsilon _{2k\uparrow }+\omega
-\varepsilon _{1k\uparrow }+i\delta \right) (\varepsilon _{2k\uparrow
}+\omega -\varepsilon _{1k^{\prime }\downarrow }+i\delta )}\right. .
\label{J2up}
\end{eqnarray}
\end{widetext}
Although the expressions seem to be long, all $J_{j\sigma}(\mathbf{q},\omega)$ terms
have the same simple structure. They contain energy-difference denominators
corresponding to the transitions from initial $j\sigma $ states to all
allowed final states, and the corresponding resonant terms. Here the single
allowed spin-conserving transition arising due to the $\mathcal{H}_{A}$ term in Eq.(\ref{eq:18})
is the momentum-conserving as well, while the two transitions caused by
random spin-orbit term $\mathcal{H}_{\mathrm{so}}^{\mathbf{k^{\prime }k}}$
are off-diagonal both in the spin and momentum subspaces, as illustrated in Fig.2. Taking the
imaginary part in each of these terms we get $\delta -$functions
corresponding to the energy conservation for the transitions to different
bands.
\end{appendix}

\end{document}